# Supermassive Binaries and Extragalactic Jets

C. Martin GASKELL

Physics & Astronomy Dept., University of Nebraska, Lincoln, NE 68504-0111, USA

**Abstract.** Some quasars show Doppler shifted broad emission line peaks. I give new statistics of the occurrence of these peaks and show that, while the most spectacular cases are in quasars with strong jets inclined to the line of sight, they are also almost as common in radio-quiet quasars. Theories of the origin of the peaks are reviewed and it is argued that the displaced peaks are most likely produced by supermassive binaries. 3C 390.3 shows precisely the change in radial velocity predicted by the supermassive binary model. The separations of the peaks in 3C 390.3-type objects are consistent with orientation-dependent "unified models" of quasar activity. If the supermassive binary model is correct, all members of "the jet set" (astrophysical objects showing jets) could be binaries.

## 1. Introduction

Many quasars show the strong jets considered in detail elsewhere in this volume. However, at a conference like this it is all too easy to forget that *many quasars do not have strong jets at all*. These quasars are the radio-quiet quasars (I will use the word "quasar" to refer to all active galaxies – see Gaskell 1987 for an introduction to quasar taxonomy). To understand how and why extragalactic jets are produced and sustained I believe it is important to understand the differences between when jets are produced and when they are not. The optical and UV spectra of quasars show strong broad emission lines coming from dense rapidly moving gas (the "broad line region"; BLR). From light-echo studies we now have a good idea of the location of the BLR (see Gaskell 1994 for a review). It extends from within a few light days of the central engine (CE) to a few light-weeks or light-months away. The size of the region depends on the luminosity (Koratkar & Gaskell 1991b). This range of radius is *precisely* the range in which jets form. It is reasonable to ask therefore whether we see differences in the BLR between radio-loud quasars (*i.e.*, quasars with strong jets) and radio-quiet ones.

There are a number of differences between radio-loud and radio-quiet quasars – line profiles, Balmer decrements, strength of the extended emission, optical Fe II emission, broad absorption lines, *etc*. Not all of these are necessarily fundamental; some differences might be consequences of differing viewing angles (see Antonucci 1993, 1996 and Wills 1996). The difference I want to focus on here is the line profile difference: quasars with strong jets, especially those not believed to be viewed face-on, have broad emission line profiles that are broader and more complex than those of radio-quiet objects (Osterbrock, Koski & Phillips 1975, 1976; Osterbrock 1977; Miley & Miller 1979). I want to explore here possible explanations of these line profiles. The explanation(s) will clearly be relevant to the question of why some quasars show strong jets while others do not.

The majority of quasars have a smooth symmetric profile that can be represented by the classic logarithmic profile (see Blumenthal & Mathews 1975, and Mathews & Capriotti 1985), except in the very core. The apparent symmetry is often illusory because the lines can be shifted relative to the rest-frame of the galaxy. This is almost always the case for the high ionization lines (Gaskell 1982) which show a 600 km s$^{-1}$ blueshift on average. Sulentic (1989) found that only ~ 15 % of a sample of quasars had Hβ profiles that were truly symmetric. In addition to these subtle asymmetries, I drew attention (Gaskell 1983b) to a class of quasars, which I will call "3C 390.3 quasars", in which the broad line peaks were displaced by *substantial* amounts relative to the rest-frame of the galaxy and in which two displaced peaks (one blueshifted and one redshifted) were sometimes present. I proposed that the displaced peaks were caused by two separate BLRs, each associated with a separate CE forming a supermassive binary (SMB) in the center of the host galaxy. I further suggested that the presence of such a binary may well be essential for the formation of collimated large-scale radio jets. In this paper I review and update the evidence for the SMB model and discuss these ideas further.

## 2. Properties of 3C 390.3 Objects

The classic example of a displaced broad-line peak object is 3C 390.3. It was the first example discovered (Sandage 1966; Lynds 1968), is by far the best studied, and shows all the features of the class. Following astronomical tradition I will call members of the class of quasars with shifted emission line peaks "3C 390.3 objects". Before discussing models of the 3C 390.3 phenomenon I first summarize the properties of the broad lines in 3C 390.3 objects, including some properties that are quite obvious.

1. *The relative velocities are large.* Relative velocities of thousands of kilometers per second are relatively common. 3C 390.3 itself has a displaced blue Balmer peak at ~ −4000 km s$^{-1}$ and a displaced red peak at ~ +2000 km s$^{-1}$. Blending of profiles makes the determination of the distribution of relative velocities difficult and small velocity differences will be missed[1].
2. *All quasars show an undisplaced peak.* Without exception, every 3C 390.3 object shows a central (undisplaced) peak. The profile of this peak is uncertain since it depends on the subtraction of the displaced components and of the narrow line region (NLR). In some cases at least, it seems to be broader than the NLR lines.
3. *A substantial part of the line flux comes from the displaced peaks.* When the peaks are well separated it is obvious that *the majority of the flux* in the Balmer lines comes from the displaced peaks (see, for example, figures 2 & 5 of Zheng, Veilleux & Grandi 1991 or figures 4*a-j* of Eracleous & Halpern 1994).
4. *Quasars with displaced peaks have broader lines than ordinary quasars.* This has been known for some time (Lynds 1968; Osterbrock, Koski & Phillips 1975, 1976; Osterbrock 1977), but I give a new analysis of this in section 4 below.
5. *3C 390.3 objects have steeper Balmer decrements.* It is well known that lobe-dominated quasars have steeper Balmer decrements than radio-quiet quasars (Grandi & Osterbrock 1978). It is therefore not surprising that radio-loud 3C 390.3 objects have steep Balmer decrements, but I show below (see section 6.7) that *radio-quiet* 3C 390.3 objects also have slightly steeper Balmer decrements.
6. *The peaks vary independently in brightness.* Like probably all quasars, 3C 390.3 is variable and the lines vary as well as the continuum. Veilleux & Zheng (1991) give a useful display of many Lick Observatory spectra. At some epochs (such as the summer of 1975) the blue displaced peak is prominent; at others (such as June 1980), the red peak is more prominent.
7. *There could always be two displaced peaks.* Not all displaced BLR peak objects show two peaks − some only show one − but I believe that if we look carefully enough or wait long enough most of the *single* displaced peak objects will show two peaks. 3C 390.3 itself is important because it shows all the characteristics of the class at various times as far as relative strengths of peaks go.

---

[1] The situation is very similar to the problem of estimating the distribution of apparent separations of visual double stars.

8. *There is no unambiguous case of more than one peak with the same sign of the displacement.* I believe that no spectra taken *at one epoch* show more than two displaced peaks or the two peaks being displaced in the same direction. I emphasize "at one epoch" because it is easy to see all sorts of peaks in difference spectra (the difference between spectra taken at different epochs). For example, if a symmetric line simply becomes broader the difference spectrum will show at least one peak in each wing. Noise bumps are also magnified in difference spectra. Peaks in difference spectra therefore need to be interpreted with great caution. The presence or absence of more than two displaced peaks is a very important issue since some models only predict two peaks and any claim of more than two peaks needs critical examination[2].
9. **The central peak varies less than the displaced peak.** If spectra of different epochs are subtracted, the central peak often subtracts out perfectly (see Wamsteker et al. 1985, figure 3, for example).
10. *The central peak has different line ratios from the displaced peaks.* This is particularly noticeable for Ly $\alpha$ which is much stronger in the central peak in 3C 390.3 (Ferland et al. 1979; Zheng, Veilleux & Grandi 1991). This is also true for non-3C 390.3 quasars (Zheng 1992).
11. *All broad lines show the displaced peak effect.* The C IV $\lambda1549$, C III] $\lambda1909$ and Mg II $\lambda2798$ lines of 3C 390.3 have profiles which are consistent with the H$\beta$ profiles of the same epoch (Perez et al 1988). The displaced blue peak is clearly visible in the C IV line, albeit shifted by ~ 900 km s$^{-1}$ (see figure 3 of Perez et al. 1988). The redward peak is harder to separate, but Perez et al. think it is also blueshifted relative to the one in H$\beta$. The 900 km s$^{-1}$ shift is the normal blueshift of high ionization lines with respect to low ionization lines (Gaskell 1982). The cause of this shift is not fully understood, but it is interesting that, at least in 3C 390.3, the separate peaks show it. It will be interesting to see whether this remains generally true as *HST* data become available for more 3C 390.3 objects. Perez et al. (1988) find other broad emission line profiles in 3C 390.3 to be basically similar to those of C IV and H$\beta$. Unfortunately all lines have blending problems that hinder detection of displaced peaks. The best line for detecting displaced peaks is probably H$\beta$, but even it is not perfect. H$\alpha$ has the [N II] lines on either side of its peak. H$\gamma$ is weak and had [O III] $\lambda4363$ on its red wing. Ly $\alpha$ is blended with N V λ1240 on the red side and tends to have absorption on the blue side. The He II lines are badly blended (He II λ1640 with C IV and [O III] λ1663; He II λ4686 with Fe II). C IV λ1549 has He II on the red wing. The higher ionization lines in quasars in general tend to be wider than the low ionization lines (Shuder 1982; Mathews & Wampler 1985) and Ly α is often con-siderably broader than the Balmer lines (Zheng 1992). This also makes detection of displaced peaks harder. The strength of the central component of Ly α noted in the previous section is a further problem for detecting displaced peaks.

---

[2] Veilleux & Zheng (1991) claim that 3C 390.3 shows a highly redshifted peak at +4600 km s$^{-1}$ in the 1974-75 Lick spectra (the main redshifted peak is at about +2000 km s$^{-1}$). However, this is almost certainly an instrumental effect caused by a spurious dip in the spectrum. This is demonstrated by the alleged +4600 km s$^{-1}$ feature being strongest in two spectra taken on the (UT) nights of June 4 and June 5, but vanishing in a longer observation with a slightly different setup the very next night. For further discussion of this, see Gaskell (1995).

In addition to these properties of the broad emission lines in 3C 390.3 quasars, there are some other differences between 3C 390.3 quasars and non-3C 390.3 quasars in their other properties:

1. *3C 390.3 objects have a larger contribution of starlight to the optical continuum.* The median percentage of starlight is about three-times greater on average in the most extreme radio-loud 3C 390.3 objects compared with other radio-loud quasars. In a number of extreme 3C 390.3 quasars the fraction is close to 100%. Eracleous & Halpern (1994) interpret this to be indicative of a weaker "blue-bump" implying a different accretion-disk structure in 3C 390.3 objects. I believe, however, that a simpler explanation, since 3C 390.3 objects also have steeper Balmer decrements, is that there is simply more reddening. This has a natural orientation explanation in unified schemes (see section 6.7 below).
2. *The equivalent widths of the forbidden lines are greater in 3C 390.3 quasars.* The median equivalent widths of the low-ionization [O I] λ6300 and [S II] λλ6716, 6731 lines of the most extreme 3C 390.3 quasars are about double those of other radio-loud quasars. In contrast, for the higher ionization [O III] λλ4959, 5007 lines, Eracleous & Halpern (1994) find no such difference in the equivalent width distributions. They conclude that the ionization parameter is systematically smaller in the extreme 3C 390.3 quasars. I believe, however, that the difference in equivalent widths can also be readily explained as an orientation effect. The low-ionization lines come from a very extended, often conical, region that will not be easily obscured from any angle. Their equivalent width is expected to increase when they are viewed edge-on. The [O III] lines, on the other hand, come from much closer in. They will suffer almost as much extinction as the continuum when the quasar is viewed close to edge-on. Their equivalent width will be much less orientation dependent. In this picture, the ionization parameter is the same on average, but the closer-in higher-ionization lines suffer greater extinction. This is supported by De Zotti & Gaskell (1984) finding broad-line

Balmer decrements to be orientation dependent while narrow-line decrements are not.

One additional alleged difference needs addressing: Halpern & Eracleous (1994) claim that the far-IR properties of 3C 390.3 quasars are significantly different from the far-IR properties of "normal" radio-galaxies. They point out that the three quasars with the flattest spectra between 25 and 60 µm "among the sample of 131 radio galaxies observed by IRAS" (Golombek, Miley & Neugebauer 1988), are all extreme 3C 390.3 quasars[3]. If this is really a significant effect, there are a number of possible explanations, but I believe there is not a significant effect. At first sight it looks as though the significance level is 1 in 9 x $10^4$, but there is a serious problem in how the comparison is made. Naturally, Eracleous & Halpern (1994) observed quasars which they initially thought would show broad lines. The IRAS sample was not so chosen. It includes "narrow-line radio galaxies", LINERs and many quasars with only stellar absorption line spectra. Broad-line quasars are in the minority. Inspection of the Golombek et al. (1988) data shows that broad-line quasars have a mean 25 to 60 µm spectral index of -0.22, while the narrow-line quasars have a mean index of -0.99, and the LINERs and non-emission-line galaxies have a mean index of -1.86. The differences between these are significant at about the 90 − 95 % level. The Eracleous & Halpern (1994) objects in common with the IRAS sample are, of course, only from the broad-line subset and it is only with these that a comparison should be made. *The 25 to 60 µm spectral indices of the 3C 390.3 quasars in the IRAS sample are not significantly different from the other broad-line quasars in the IRAS sample*. In other words, I believe the 3C 390.3 objects have flat far-IR spectra *because they are broad line objects*, not because they are 3C 390.3 objects. A preliminary check of some of the 25 to 60 µm spectral indices of radio-*quiet* 3C 390.3 objects shows that they are similar to the radio-loud ones and that, again, there is no statistically significant difference between these objects and the "non-3C 390.3 type" ones.

Why does the far-IR emission depend on whether we can readily see the BLR? Since far-IR emission is essentially completely isotropic the explanation cannot be in orientation effects. Miley et al. (1985) and Golombek et al. (1988) argue that a combination of two dust components is the likeliest interpretation of the radio galaxy data. The main body of the galaxy produces the coldest emission while the nuclear region has warmer dust that raises the 25 µm flux. They do not point out the difference between broad-line and narrow-line quasars, but I believe it can be explained as follows: The main variable is the amount of dust in the host galaxy. This dust produces the 60 µm or longer wavelength emission; the more cool dust there is the steeper the far-IR spectrum. A lot of dust also makes it less likely that we will be able to see BLR; even more dust will make it impossible even to see the NLR.

---

[3] Actually, "131 radio galaxies" is misleading. Although Golombek et al. (1988) observed this number, they only *detected* 58 of them.

## 3. Frequency of Occurrence of 3C 390.3 Objects

Gaskell (1983b) noted that the best cases of displaced broad lines were in extended radio sources (= "lobe-dominated" quasars). This still remains true, but the availability of high-quality profiles has revealed that many radio-*quiet* quasars also have displaced peaks. Osterbrock & Shuder (1982) and Stirpe (1990) give good samples of profiles. These are particularly valuable because of the careful subtraction of narrow lines, and in case of Stirpe (1990), of Fe II emission and the stellar continuum as well. I have visually classified the profiles in the Stirpe atlas into "no displaced peak", "single displaced peak" and "double displaced peak". If a displaced peak has a small displacement it will appear only as an inflection on the side of the central peak unless it is very strong. If it is very strong, the line peak will be shifted from the systemic redshift of the host galaxy. I therefore identified such inflections as displaced peaks. I summarize the results in the table 1 (the Osterbrock & Shuder 1982 spectra are consistent with these results).

**Table 1.** Statistics of displaced emission lines in radio-quiet quasars.

| Profile Type | Percentage |
|---|---|
| No obvious displaced peaks | 40% |
| Single displaced peak | 25% |
| Two displaced peaks | 35% |
| | |
| Blue peak strongest | 29% |
| Red peak strongest | 25% |
| Peaks approximately equal | 6% |

The most striking thing in this table is the high percentage of displaced peaks. This is particularly so since the percentage of displaced peaks must have been underestimated, possibly by a lot. Displaced peaks vary. For example, Mrk 6 (not a member of the Stirpe sample) shows no displaced peak when it is in a low state, but when it is bright it has a clear single blueward-displaced peak (see Rosenblatt et al. 1994, figure 14). The percentages of displaced peaks would also certainly be slightly higher still if my search had been conducted by subtracting off an appropriate central peak or subtracting spectra taken at different epochs. Likewise, such analyses would reveal more second peaks in the objects that are currently classified as only having single displaced peaks. The other thing to notice from table 1 is that the percentages of blue and red peaks are about the same. The excess of blue peaks is not statistically significant, but models invoking organized bulk motions to explain the displaced peaks predict that the blue peaks will be slightly stronger because of relativistic effects (see Mathews 1982).

Eracleous & Halpern (1994) have undertaken a useful large survey of radio-loud quasars. Their motive was to search for quasars with emission lines that might have arisen from a disk, but their sample can be used to look at the statistics of displaced emission line peaks. Unfortunately, unlike Stirpe (1992), they do not report H$\beta$ as well, nor do they subtract the narrow emission lines (the [N II] lines are particularly bothersome), and they only show starlight subtracted profiles for quasars which they considered "disk-like". This makes their survey a little less sensitive to displaced peaks than the Stirpe atlas. I classified the Eracleous & Halpern spectra in the same way as the Stirpe spectra. I excluded a few spectra from the sample because they were not classifiable. This was either because of the strength of [N II] emission, the narrowness of H$\alpha$, or a poor signal to noise ratio. The remaining spectra give the following statistics:

**Table 2.** Statistics of displaced emission lines in radio-loud quasars.

| Profile Type | Percentage |
| --- | --- |
| No obvious displaced peaks | 30% |
| Single displaced peak | 32% |
| Two displaced peaks | 38% |
| Blue peak strongest | 26% |
| Red peak strongest | 28% |
| Peaks approximately equal | 16% |

The first thing to notice from this table is the higher percentage of displaced peaks. This is particularly significant in light of the caveats just given. What is not clear from table 2 is that the displaced peaks are also more obvious and spectacular. Although *most* of the spectacular cases of displaced broad-line peak objects are strong extended radio sources, *not all are* (this contrary to what Eracleous & Halpern 1994 state). IC 4329A is a clear counter-example (Disney 1973). IC 4329A has a strong red displaced peak at a $\Delta z$ of +2000 km s$^{-1}$, but IC 4329A is not a strong radio source[4]. The second thing to notice from table 2 is that apart from the higher overall percentages, the ratio of occurrence of blue to red peaks is again approximately equal[5]. As was noted for the radio-quiet objects, the percentages of displaced peaks given in table 2 *must be underestimates* because of variability. Pictor A is a good illustration of this. If it were classified using the 1983 spectrum of Filippenko (1985), it would be classified as having no obvious displaced peaks, but the 1994 spectrum taken by Halpern and Eracleous (1994) shows clear double displaced peaks. It is important to remember, when considering profile classification, that essentially all quasars vary and the prototypical object, 3C 390.3 itself, could be placed in *any* of the categories in table 2 depending on when the observation was made!

3C 390.3 objects cover a range in optical luminosity running all the way up to high-luminosity objects. OX 169 with $M_{abs} = -24.7$ (for $H_o = 50$ km s$^{-1}$ Mpc$^{-1}$) is a high-luminosity example (Gaskell 1981). A couple of selection effects are expected to be working against detecting high-luminosity 3C 390.3 objects. Because they are rare, high-luminosity quasars are not seen nearby. They therefore have high redshifts and only the rest-frame UV lines are readily seen. The difficulties in detecting displaced peaks in the UV lines (see 2.11) make the detection of high-redshift 3C 390.3 objects difficult, but they do exist. PKS 0119-046 (z = 1.953, Mabs = $-28.5$) has a narrow component to Ly $\alpha$. Ly $\alpha$ and C IV have broad peaks blueshifted by $-3000$ km s$^{-1}$ relative to the narrow peaks (Gaskell 1983a). This *could* be an extreme example of the high ionization blueshifting effect (Gaskell 1982), but the relative velocity is more consistent with PKS 0119–046 being a 3C 390.3 object.

Although 3C 390.3 objects tend to be lobe-dominated quasars, there are exceptions. OX 169, for example, is core dominated. As Eracleous & Halpern (1994) point out, 3C 390.3 objects fall into both Fanaroff & Riley classes and they "have no special characteristics in the radio-band that would distinguish them from typical broad-line radio galaxies."

---

[4] IC 4329A does however have a compact core with a luminosity that places it at the upper end of the Seyfert galaxy radio luminosity function (Giuricin et al. 1990a).
[5] Eracleous & Halpern (1994) themselves identify a lower number of displaced peaks, because their motivation is fitting them with their disk model. Eracleous & Halpern mention an excess of dominant blue peaks but this is only the case when only the double displaced peak objects are considered. If we consider the single displaced peak objects as well, the numbers of red and blue dominant peaks are the same.

# 4. Widths of Emission Lines

3C 390.3 objects have long been noted for their very broad lines (e.g., Lynds 1968; Osterbrock, Koski & Phillips 1975, 1976). It is also well known that lobe-dominated radio-loud quasars have broader lines than core-dominated or radio quiet ones (Miley & Miller 1979). There is good evidence that this is an orientation effect (Wills & Browne 1986). However, I want to show here that the width of lines in 3C 390.3 objects is not just a result of their predilection for radio-loud quasars. This can be seen if we look at the median line widths of H$\alpha$ in the two radio-loud and radio-quiet quasar samples considered above (see table 3). From table 3 it is obvious that *regardless of whether there are strong radio jets or not*, the displaced-peak quasars have wider lines.

Note the progression from no displaced peak to single displaced peak to double peaks. It is important to note that most of the differences in line width can be accounted for *by the velocity shifts of the displaced peaks themselves*. For the radio-quiet objects a typical shift is about 1250 km s$^{-1}$. Thus the FWHM goes from 2150 km s$^{-1}$ when no peak is present to 2150 + 1250 = 3700 km s$^{-1}$ when one displaced peak is seen and to 2150 + 1250 + 1250 = 4650 km s$^{-1}$ when two are seen. The separations are larger for the broader line objects. Average increases of 2650 km s$^{-1}$ will similarly explain increases in both the FWHM and FWZI in the radio-loud sample.

**Table 3.** Median FWHM and FWZI (km s$^{-1}$) of H$\alpha$ for different profile types.

|  | Radio-quiet FWHM | Radio-loud FWHM | Radio-loud FWZI |
|---|---|---|---|
| No obvious displaced peak | 2150 | 4300 | 17800 |
| Single displaced peak | 3400 | 5000 | 19800 |
| Double displaced peaks | 4700 | 9800 | 12900 |

# 5. Models

In this section I review the main models proposed to explain 3C 390.3 objects. But first, one consensus should be noted: everybody agrees that the central peak is due to a separate component of gas that is probably somewhat further out than the gas producing the displaced peaks. This is because of the differing line ratios and lack of variability. I believe that this gas is the "intermediate line region" of Wills et al. (1993) and Brotherton et al. (1994). I discuss this further in section 6.8.

## 5.1 Obscuration Effects

Capriotti et al. (1979) and Ferland, Netzer & Shields (1979) suggested very asymmetric profiles were the result of obscuration and radial motion. Although good fits can be obtained to some lines (see Capriotti et al. 1979) the problem with this explanation is that blue and red asymmetries are equally prevalent (see section 3). The model is in the unsatisfactory situation of requiring BLR inflow in some objects but outflow in others (Gaskell 1983b).

## 5.2 Self-Absorption

A number of quasars show intrinsic absorption line systems with redshifts close to the emission-line redshift (e.g., NGC 4151; Anderson 1974). A number of people suggested that the dips in the profiles of 3C 390.3 objects are due to self-absorption (e.g., Boksenberg & Netzer 1977; Smith 1980). UV studies have shown that when there *is* such absorption, it is quite narrow and does not produce the broad gentle dips seen in 3C 390.3 objects. Also, self-absorption is usually on the blue side of an emission line so it is an unlikely explanation of the very redshifted displaced peaks. Nonetheless, the role of self-absorption should not be forgotten since it does occur to some degree (see, for example, the H$\beta$ and C IV profiles in NGC 3516; Wanders et al. 1993)

## 5.3 Light Echo Effects

Capriotti et al. (1982) suggested that displaced BLR peaks were the result of pulses of the photoionizing continuum in an ordered BLR velocity field. While such a model can fit some lines, the light travel times across BLRs are now known to be much too small by more than an order of magnitude (Gaskell & Sparke 1986) for this idea to be tenable. The wavelengths of displaced peaks do not vary on a timescale as short as the light-crossing timescale (Gaskell 1983b).

## 5.4 Ejection of Clumps of BLR Material from the Central Engine

This was the first explanation of displaced peaks offered (Burbidge & Burbidge 1972; Osterbrock, Koski & Phillips 1975). The idea is attractive because radio-emitting plasma is clearly ejected in the jets and we see broad blueshifted UV absorption lines in some radio-quiet quasars. Bi-conical ejection was suggested for 3C 390.3 itself by Oke (1987) and explored in detail by Zheng, Binette & Sulentic (1990) and Zheng, Veilleux & Grandi (1991). The latter were able to obtain good fits to a couple of epochs[6]. The main problem with the jet model (Oke 1987, Gaskell 1988b) is that it is the *opposite* of what is expected in unified models. Core-dominated sources (where the jet is coming towards us) would be expected always to show strong displaced peaks while lobe-dominated sources where the jet is at an angle to the line of sight would show smaller shifts of the peaks. As we have noted above, some of the most spectacular cases are in lobe-dominated quasars. Unified models therefore imply that the bulk motions producing displaced peaks are in the plane *perpendicular* to the jet axis.

Unfortunately, a large number of variability studies of quasars show that the net motion of BLR gas is *not* a simple outflow (Gaskell 1988a, Koratkar & Gaskell 1989, 1991a, Crenshaw & Blackwell 1990, Maoz et al. 1991, Korista et al. 1995)[7]. These studies all show that in the majority of cases, if anything, there is a slight net *inflow*, because the redshifted C IV wing tends to lead the blue wing slightly (see also Done & Krolik 1996).

Bi-conical outflow models also have difficulty explaining why there are so many *single* displaced peak objects (especially when the peak is on the red side).

## 5.5 Anisotropic Illumination of the BLR Gas

Zheng, Veilleux & Grandi (1991) also mention the possibility that double peaks could be produced by anisotropic illumination of the BLR (a "searchlight"). This also requires net inflow or outflow of the BLR. There is now some evidence that the ionizing radiation is confined to a wide conical beam as it illuminates the BLR in at least one object (NGC 5548; Wanders et al. 1995). Goad et al (1996) show that double-peaked profiles arise when the observer lies outside this cone. The further the observer is outside the cone the more double peaked the lines appear and the weaker the continuum becomes. It is not clear yet whether a satisfactory fit to the profiles of the extreme 3C 390.3 objects can be obtained with parameters that are consistent with the observed slight inflows (since the variability studies imply that the *dominant* motion is not pure inflow). The model also has to explain the smooth shift in wavelength of the blue peak in 3C 390.3 (see figure 1 below).

---

[6] However, everybody always gets good fits to the line profiles! The ability to reproduce the profiles is unfortunately not a way to distinguish between models.

[7] For a possible way of reconciling these observations with outflow see Kundt (1988).

**5.6 Accretion Disk Models**

Since quasars often show a clear axis of symmetry (the line the jets emerge along) something in the inner region is rotating, probably the central engine itself. Not surprisingly, as Mathews (1982) puts it, "Rotational energy production and accretion disks as explanations of broad emission lines have been regarded as possibilities since the earliest quasar literature (Woltjer 1959; Lynden-Bell 1969)." He gives an already quite long list of papers up to 1981 that give models employing rotational broadening and mentions that there are "numerous other studies" in which favorable allusions to the possibility appear. Now, 14 years later, the list would be many times longer and I will not attempt to give one. The interested reader is referred to Collin-Souffrin et al. (1988), Collin & Dumont (1989), Dumont & Joly 1992 and Rokaki (1994) for theoretical arguments in favor of line emission from disks and further references. Mathews (1982) calculates theoretical profiles from disks and concludes for a variety of reasons that "rotation is an extremely unlikely means of producing broad emission lines". One of Mathews's biggest arguments is that quasar emission line profiles do not resemble those expected from disks. The realization about five years later that quite a number of quasars actually *do* show double-peaked profiles, or at least could be made to show double profiles by suitable differencing of spectra led many authors to propose that the displaced peaks were produced in disks (Oke 1987, Alloin, Boisson & Pelat 1988, Stirpe, de Bruyn & van Groningen 1988, Halpern & Filippenko 1988 and Perez et al. 1988). There have subsequently been a number of attempts to fit disk models to observations. The largest attempt has been by Eracleous & Halpern (1994).

Gaskell (1988b) identified three problems with disk models:
1. Although fits can be obtained in *some* cases[8], the ratio of peak intensities is often not right (this point is also made by Sulentic et al. 1990). Usually only one displaced peak is prominent, and in many cases the other cannot even be detected (see tables 1 and 2 above). Also, about 50% of the time it is the *red* peak which is strongest. As Eracleous & Halpern (1994) concede, these objects cannot be explained by the simple disk model. These are a large fraction of the displaced peak quasars. Eracleous & Halpern suggest that the SMB model might be appropriate for some of these.
2. There is no natural explanation of the "normal" ("classic" or "logarithmic") line profile in the disk model.
3. Line variability rules out disks. This is the strongest argument against disks. The continuum of quasars is seen to vary and the lines vary in response to the continuum changes (this is the whole basis of the quasar reverberation mapping cottage industry!). If the lines arise in a disk, and are ionized by a central source, then *both peaks must always vary up and down together*. In general they do not (Oke 1987 - see his figure 1; Peterson, Korista & Cota 1987; Miller & Peterson 1990; Veilleux & Zheng 1991 - see their figure 2*e*)[9]. A disk model which fits at one epoch will not fit at another epoch! Even if a disk model can be made to fit at more than one epoch, an unphysical change of parameters might be needed (e.g., in the inclination of the disk).

---
[8] See footnote 6.

In order to circumvent the first of these problems Zheng, Veilleux & Grandi (1991) modified the simple disk model to include a "hot spot". There is precedence from cataclysmic variable stars for including such hot spots (see Zheng et al. 1991). They obtain satisfactory fits to two epochs of 3C 390.3 in this manner[10]. In a similar vein, Eracleous et al. (1995) introduced highly elliptical disks to get round problem number one. They are able to fit some of their profiles that could not be fit with simple circular disks[11], but I think that both this model, and the Zheng et al (1991) one, are probably already excluded by existing variability observations.

There are a couple of additional problems for disk models:
1. All disk models need to introduce an arbitrary broadening by some additional mechanism.
2. Although the fits to the two humps can be acceptable, the disk models do not fit the high velocity wings, as Eracleous & Halpern (1994) point out (this is despite having 7 free parameters in the circular disk model and 9 in the elliptical disk model).
3. The disk model makes very specific predictions of the percentage polarization of the broad lines, the polarization position angle within the line, and the shape of the polarized flux (Chen & Halpern 1990). Unfortunately observations do not confirm these specific predictions (Antonucci, Hurt & Agol 1995). The polarization observations do, however, show that the wings are polarized and probably do arise in a region separate from the line core (but this is the one point that almost all models agree upon!)
4. It is difficult to get a disk to emit lines. Most workers on this model favor an external source of ionization shining or reflecting back onto the disk.

---

[9] Some authors *have* claimed that some profile variations are consistent with a disk origin however (Alloin, Boisson & Pellat 1988; Stirpe, de Bruyn & van Groningen 1988, but see also Marziani, Calvani & Sulentic 1992).
[10] See footnote 6
[11] Again, see footnote 6!

## 5.7 Supermassive Binary (SMB) Model

Gaskell (1983b) suggested that the displaced peaks in 3C 390.3 objects were due to two separate BLRs each with its associated central engine. The associated central engines are necessary because the BLRs will not survive without them. The high displacement velocities of several thousand km s$^{-1}$ are the projected velocities as the engines orbit each other. The idea that quasars consist of some sort of binary had been originally proposed by Komberg (1968). In the late 1970's the discovery of jets from the galactic stellar-mass binary system SS 433 followed by radio observations suggesting that the radio axes of some quasars wobble or precess (Ekers et al. 1978) led to further consideration of the idea (Collins 1980; Begelman, Blandford, & Rees 1980; Whitmire & Matese 1981).

There are several ways a supermassive binary in the center of a galaxy could form, but the most sure-fire one is through mergers (Blandford, Begelman & Rees 1980; Roos 1981). Galaxy mergers are very common and still going on today. A large fraction of galaxies, perhaps 50%, show at least low-level quasar activity, (see Keel 1985 for a review), so some sort of central engine must be present. Also, even when there is no quasar activity, we see signs of inactive supermassive objects (e.g., in M 31 and M 32 in the local group). Therefore many mergers *must* form supermassive binaries. As two galaxies merge, the two nuclei will spiral together. This is beautifully shown in the video of Barnes (1993). It is important to realize that this is something common that is happening all the time. There can be no doubt that supermassive binaries form. Indeed, if 3C 390.3 objects eventually prove *not* to be a consequence of the formation of supermassive binaries, it is reasonable to ask, "what *are* the consequences?"

The SMB model correctly predicts several things:
1. *The BLR profiles*. Because of its non-uniqueness, profile fitting does not "prove" the correctness of a model, but displaced peaks in difference-spectra, where one hopes that the central stationary component has cancelled out, can be fit well with just a pair of Gaussians (six parameters). Good examples can be seen in Akn 120 (Alloin, Boisson & Pelat 1988) and Pictor A (Halpern & Filippenko 1994). In both of these cases, a disk model is a poor-fit, even for $i = 90^o$. This is because the disk models do not reproduce the lack of emission between the peaks. For

Pictor A, the two Gaussian-like components in the difference spectrum are well separated. Roughly equal peaks, as in the case of Pictor A are not the rule. The red peak in the Alloin, Boisson & Pelat (1988) difference spectrum of Akn 120 is 50% stronger than the blue peak. The Wamsteker et al. (1985) difference spectrum of Fairall 9 in the early 1980's is flat apart from one approximately Gaussian, well-defined, red displaced peak. The NGC 5548 difference spectra from the mid-1980's are flat apart from a similarly well-defined peak to the blue (Peterson, Korista & Cota 1987).

2. *The velocities seen.* An SMB is most likely only to be seen in a limited velocity range because there are natural lower and upper limits to the orbital velocity (Gaskell 1983b). The initial merger takes place on the dynamical time-scale and the orbit decays fairly rapidly because of tidal friction (Begelman, Blandford & Rees 1980). Again, this is well illustrated in the Barnes (1993) video. Although we do see a few cases of mergers of quasars in progress (such as Abell 400, Mrk 78 and Mrk 266), we expect them to be rare, simply because they happen so quickly. As the orbital velocity of the binary goes up, the Rutherford gravitational cross-section goes down and the rate of evolution decreases (Begelman, Blandford & Rees 1980). The timescale will increase when the orbital velocity exceeds the stellar velocity dispersion (a few hundred km s$^{-1}$), so this is a *natural lower limit* to the orbital velocity of observed SMBs (Gaskell 1983b). At high velocities, gravitational radiation causes rapid decay of the orbit when the speed starts to become relativistic (Begelman, Blandford & Rees 1980). This causes a *natural upper limit* to the orbital velocity of a few percent of the speed of light (Gaskell 1983b).

3. *There are only two displaced peaks at most* (Gaskell 1983b). Multiple mergers are possible, but three-body systems are unstable except in certain particular cases. One or more components will be ejected from the galaxy via the "gravitational slingshot" mechanism (Saslaw, Valtonen & Aarseth 1974; Makino & Ebisuzaki 1994). Any stable hierarchical systems (supermassive analogs of α Geminorum *etc.*) which do form will have relatively short lifetimes because of gravitational radiation.

4. *The peaks can be of very different brightness.* Since the luminosity of the BLR correlates well with the luminosity of the central engine (Yee 1980; Shuder 1981), and the luminosity of the central engine correlates with its mass (Koratkar & Gaskell 1991b and references therein), the model predicts that when the average luminosities of the two components are unequal, the brightest component will have the smallest velocity shift.

5. *The two peaks will vary independently in brightness* because they have independent continuum sources. This seems to be the case in every 3C 390.3 object seen to vary so far (Wamsteker et al. 1985; Peterson, Korista & Cota 1987; Miller & Peterson 1990; Veilleux & Zheng 1991; Marziani et al. 1993). Each BLR will vary with its continuum. We will see the combined effects. The most variable component, in the absolute sense (which is probably also the brightest one) will dominate the observed continuum variability. Only one of the two peaks will be observed to follow the continuum. This is obviously the case for 3C 390.3 (Veilleux & Zheng 1991)[12] and has also been claimed for NGC 5548 (Peterson, Korista & Cota 1987).

6. *There will be smooth radial velocity changes.* Gaskell (1983) estimated that orbital periods would be of the order of centuries and therefore that radial velocity changes could be detectable in a few decades. Although Halpern & Filippenko (1988) claimed that the lack of a radial velocity change in Arp 102B ruled out the SMB model, Veilleux & Zheng (1991) reported what seemed to be a shift in the wavelength of the blue peak in 3C 390.3 in Lick Observatory spectra taken by D.E. Osterbrock, J.S. Miller and their collaborators. Marziani et al. (1993) report a variation of the displacement of the peak of Hβ in OQ 208 *with luminosity* (rather than with time). I believe this is not a real effect but the result of the blending of two components. My analysis of over two decades of 3C 390.3 data (see figure 1) shows that there is indeed a very clear and convincing radial velocity curve (Gaskell 1995). This is compatible with the period of a few centuries predicted by Gaskell (1983b). The weakness of the Filippenko & Halpern (1988) argument is the assumed total mass.

# 6. Discussion

### 6.1 System Parameters for 3C 390.3

The radial-velocity curve for 3C 390.3 (Gaskell 1995 − see figure 1) allows the derivation of a number of parameters. The minimum orbital period that provides an acceptable fit to the curve is 210 yr, with 300 yr being more likely ($\chi^2 = 0.8$ per degree of freedom; see Gaskell 1995 for details). This gives a projected maximum velocity ($v \sin i$) of 5340 km s$^{-1}$. It is probably safe to assume that the orbits have been circularized. The synchrotron self-Compton model and relativistic beaming considerations give an estimate of $i$, the inclination of the 3C 390.3 jets to the line of sight (Ghisellini et al. 1993). The ratio of optical to radio core flux supports such estimates (Wills & Brotherton 1995; Wills 1996). For 3C 390.3, Ghisellini et al. get $i \sim 29°$. This gives a true orbital velocity of 11,000 km s$^{-1}$. The minimum radius of the orbit is therefore at least one light-year. The ratio of displacements of the peaks is 2:1 so the minimum masses of the two central engines are 4.4 x $10^9$ and 2.2 x $10^9$ solar masses. These are slightly higher than the mass predicted by the Koratkar & Gaskell (1991b) mass-luminosity relationship, but in good agreement with the more reliable estimate of 2.4 x $10^9$ solar masses, from HST observations, for the mass of the CE in M 87 (Ford et al. 1994). They are also consistent with masses of remnants deduced from quasar counts (Soltan 1982) and the masses of $10^9$ solar masses for the dark central objects in M 104 (Kormendy 1988) and NGC 3115 (Kormendy & Richstone 1992).

---

[12] The plot of the *ratio* of fluxes of the blue and red peaks versus time looks smoother than either individual light curve. I think the smoothness is because the errors of calibrating the fluxes in the standard way with respect to the narrow-line region lines cancel out when the ratio is taken.

**Fig. 1** Radial-velocity curve for the blue displaced broad peak of Hβ in 3C 390.3. The curves are for periods of 210 yr, 300 yr and an infinite period (from Gaskell 1995).

The spin axes of the central engines will undergo geodetic precession. We can derive the minimum precession period from the derived masses and orbital size. Equation 8 of Begelman, Blandford & Rees (1980) gives a period of 4 x $10^5$ yrs.

For a given orbital velocity, the radius of the orbit and the period will both vary linearly with the sum of the masses. On the other hand, for orbits of approximately the same size, the period will increase with the square root of the mass while the orbital velocity will decrease with the square root of the mass. Since the observed separations of peaks are similar, we would expect the less luminous sources to show the most rapid radial velocity changes, if they do indeed have lower central-engine masses.

Note that the size and mass deduced are the same in a disk plus hot-spot model.

## 6.2 BLR Kinematics

As was pointed out by the late Michael V. Penston, if BLR cloud motions are gravity dominated, a supermassive binary does not give a line profile that is simply the sum of the profiles around two separate central engines. This is because the low-velocity gas in the peaks would have to arise from far out (see footnote 3 of Cheng, Halpern & Filippenko 1989). If the SMB model is correct, it has important implications for BLR cloud motions: they have to be sub-virial. Since non-gravitational forces such as radiation pressure and winds are already known to be significant (indeed, they are the dominant force for broad absorption line clouds in quasars), this is not unreasonable. If we ignore drag forces, ignore the extended mass distribution of the stars, and consider only radiation pressure in addition to gravity, the clouds will be subject to a net inverse-square law force which is less than the force of gravity. The velocities at a given distance will therefore be lower than the virial velocities. The masses Koratkar & Gaskell (1991b) derived from BLR cloud motions will be too small. Mathews (1993) has gone further and included the effects of drag forces. He shows that masses have been underestimated by factors of ~ 10–20.

## 6.3 Implications for Reverberation Mapping

The separation deduced for the components of 3C 390.3 is not much greater than the BLR sizes in comparable objects (Koratkar & Gaskell 1991a,b; see Peterson 1994 for a recent review of the many new results), so, while each component of the BLR will be dominated by the variations of its corresponding central engine, the light from the brightest one may well strongly influence the BLR of the other. Thus, when a 3C 390.3 object goes into a "low state" (*i.e.*, the brightest continuum source fades), *both* BLR components could fade, but the time delay will be different. The behavior of 3C 390.3, as reported by Veilleux & Zheng (1991), is consistent with this.

## 6.3 Jet Structure

Curving radio jets, the motivation of Begelman, Blandford & Rees (1980) to propose SMBs, continue to provide strong support for the idea (Hunstead et al 1984; Roos 1988; Roos, Kaastra & Hummel 1993; Conway & Murphy

1993; Conway & Wrobel 1995). Any reader in doubt about precessing radio jets in quasars is referred to the beautiful radio maps of the PKS 2300-189 jets (Hunstead et al. 1984)[13]. To quote Begelman, Blandford & Rees, "...the massive binary seems to be the only way to set this [quasar jet precession] in a dynamical context". Precessing quasar jets are common. Hutchings, Price & Gower (1988) estimate that 30% of jet sources show evidence for precession. There is also at least one good example of helical *optical* emission, NGC 3516 (Veilleux, Tully & Bland-Hawthorn 1993), which also has a clear single displaced BLR emission-line peak (Wanders et al. 1993).

The determination of a precession period from jet structure is difficult, but there are many estimates in the literature. These are derived by fitting precessing jet models to helical structure in radio maps. References for more than a dozen can be found in Roos (1988) and Lu (1990). They find a correlation between the precession period and the radio and optical luminosities. I believe, however, that the selection effects in this need careful consideration.

Begelman, Blandford & Rees (1980) give formulae for the orbital and precession periods in the SMB model. As noted above, the minimum precession period for 3C 390.3, based on the parameters derived from the optical observations, is typical of precession periods deduced from radio maps. The precession periods are, of course, much longer than orbital periods. It is interesting that the shortest *precession* periods, as deduced from radio maps (for 3C 273 and 3C 345) are ~ $10^3$ yrs − only slightly longer than the *orbital* period I get for 3C 390.3 from its radial-velocity curve. If these short precession periods are correct, then *very* rapid orbital motion might be seen in the optical spectra − perhaps on a timescale of only a year. This, and the usual temporal undersampling of quasar spectra, could produce some of the changes seen in some line profiles.

Before too much comparison of the radio and optical results is carried out, one should note that some caution is necessary in comparing samples. This is because different selection effects are at work in the two cases. Only motion with a long period produces obvious changes in jets. Only rapid motion produces readily separated line peaks in optical spectra. Precession is most obvious in jets close to our line of sight; orbital motion will be most obvious when jets are in the plane of the sky.

### 6.4 VLBI Cores

At least *some* SMBs should produce double radio cores. It should be remembered that the majority of quasars are *not* radio-loud, so the chances are that only *one* of the central engines will be a strong radio source. The unambiguous cases of double radio cores have fairly wide projected separations: Abell 400 cD = 3C 75 (Owen et al. 1985) and Mrk 266 (Mazzarella et al. 1988) both have separations of 8 kpc; NGC 3256 has a projected separation of 1.4 kpc. The nuclei in these sources are also roughly equal in luminosity, so some selection effect is going on. There will be many more cases of *unequal* radio luminosities. These are either lost because of limited dynamic range or written off as "background" sources. The three double cores just mentioned must represent the early stages of mergers. By contrast, the binaries selected by optical spectroscopy will have separations of 1 pc (or less, for the lowest mass cases). Some of these could possibly be detected with VLBI.

---
[13] Optical spectra of PKS 2300-189 show double displaced peaks.

Double VLBI cores should be fairly rare. Radio-loud quasars are a minority of quasars. Radio core emission is also believed to be highly anisotropic because of relativistic beaming. We choose to observe radio cores because at least one of the central engines is radio-loud. Even if *both* members of the binary are radio-loud, chances are that the other one will not be beamed in our direction. Since VLBI maps have limited dynamic range, many weak emission features currently written-off as "noise" could prove to be second central engines. Radio-loudness probably also only has a limited lifetime, so both central engines might not be radio-loud at the same time.

The "classic" symmetric parsec-scale double with equal strength "mini-lobes" (see Conway et al. 1994 for a recent description of parsec-scale doubles) almost certainly has nothing to do with SMBs. Most of the asymmetric ones are probably also unrelated. However, a binary radio core is one possible explanation of the unusual VLBI structure of the displaced BLR peak quasar 4C 39.25. VLBI maps of 4C 39.25 show both stationary components *and* superluminal motion (Shaffer et al. 1987, Alberdi et al. 1993).

There is no case known of *active* double kpc-scale jets coming from separate cores with parsec-scale separation (the beautiful pairs of jets in 3C 75 start at least 8 kpc apart). There are several possible explanations of this. Perhaps they exist but are rare and we have simply not found or recognized a case yet. Perhaps the separate jets merge into one under the guiding of the combined magnetic field structure. The existence of misdirected "fossil jets" (see next section) suggests that perhaps pairs of jets from double cores are rare and that we might have just missed them.

**6.5 Fossil Jets**

In their maps of the kpc-scale structure of some bright FR II sources, Black et al. (1992) point out some cases where there have been "drastic changes of orientation in the past $10^8$ yrs." 3C 223.1 and 3C 403 are good examples. These remarkable changes of orientation are not possible from a single massive rotating object. There are, however, two possible ways of getting such changes in the SMB model. The first is for the two components in an SMB to have differently oriented rotation axes. The first one emits a jet along its axis for some period of time and then the other emits a jet much later. The second possibility is for the jet to be emitted by just the most massive engine in the binary. When the members of the binary finally coalesce there can be an abrupt change in rotation axis.

### 6.6 Mergers

It has been know for a long time that quasar host galaxies have undergone interactions and mergers (see, for example, the review by Hutchings, 1983). Since SMBs probably arise from mergers, should we expect to see more signs of interactions around 3C 390.3 objects compared with non-3C 390.3 quasars? This is a question worth investigating. It is already known that Seyferts which show greater signs of interactions have stronger radio-cores and total radio-power than Seyferts showing less signs of interaction (Giuricin et al. 1990b). In doing this sort of comparison it should be remembered that SMBs can last a Hubble time. A recent interaction might only have recently fueled an SMB that actually formed much earlier.

Mere interaction of systems alone enhances activity, as does the presence of a close companion. Some quasars are isolated and undisturbed. Even in these cases, however, a merger might have taken place *with a small companion galaxy* (Gaskell 1985). This is very possible because of the shear numbers of such galaxies. Since dwarf galaxies can also have massive central objects (*e.g.*, M32 - Tonry 1987), even an encounter with a small galaxy could produce a SMB.

### 6.7 Orientation Effects

The ratio of radio-core to radio-lobe flux density (*R*) is a well-established indicator of the orientation of the spin axis of a quasar's central engine to the line of sight (see Antonucci 1993, 1996; Wills 1996). Wills & Brotherton (1995) have shown that the ratio of fluxes from the radio and optical cores ($R_V$) is an even better indicator of orientation (see also Wills 1996). Wills & Browne (1986) established that there is a strong correlation between line width and *R*. The correlations with $R_V$ are even stronger (Brotherton 1995). These correlations imply that BLR motions are perpendicular to the jet axis. Wills & Browne (1986) interpret this in terms of a disk model of the BLR but it is also consistent with SMBs being responsible for the extra broadening in the non-face-on quasars. Indeed *the Wills & Browne*

*correlation might be mostly due to the orbital motion of the SMBs.* Whether this is true or not, clearly, the deficiency of 3C 390.3 objects in core-dominated sources is an orientation effect. As already discussed in section 2, the differences in continuum and NLR properties are naturally explained as orientation effects.

While the unified picture of core-dominated versus lobe-dominated quasars is well accepted, there is no such agreement over why some quasars are radio-quiet. Nonetheless, orientation effects also play a role in the optical properties of radio-quiet quasars. Seyfert 1 galaxies (low-luminosity radio-quiet quasars showing a BLR) are preferentially seen face-on, while Seyfert 2 galaxies (ones where only the extended NLR is seen) are not (Keel 1980). There is good evidence that at least some Seyfert 2's are quasars seen from the side. Imaging, spectropolarimetry and statistics show that there is an "ionization-cone" with an opening angle of $30 - 60^o$ (see Antonucci 1993). Because of the selection bias towards objects with a strong UV excess, most radio-quiet quasars are probably ones oriented so that the observer is inside that cone. This explains the similarities between the optical spectra of *core-dominated* radio-loud quasars and UV-excess-selected radio quiet quasars (Baldwin, Wampler & Gaskell 1989). I hypothesize therefore that *the reason why the displacements of peaks in radio-quiet quasars are smaller than in radio-loud quasars is primarily because of orientation effects*, and is not something fundamental related to whether a quasar has a strong jet or not (this is the opposite of what I proposed in Gaskell 1983b). Under this hypothesis, we see the best cases of 3C 390.3 objects in lobe-dominated quasars because extended radio lobes are a good way of finding quasars that do not have their radio jets aimed at us.

Testing this hypothesis will be tricky since we do not have a convenient orientation indicator like $R$ or $R_V$ available for radio-quiet quasars. The hypothesis does lead to four predictions about where the most spectacular cases of 3C 390.3 objects will be found:

1. *They will be found among quasars with the broadest lines.* This has already been shown to be the case (see table 3).
2. *They will be found among quasars with edge-on host galaxies.* IC 4329A provides support for this hypothesis since it is one of the most edge-on active galaxies known (De Zotti & Gaskell 1984).
3. *They will be found among quasars with a steep Balmer decrement.* The Balmer decrement correlates with host galaxy orientation (De Zotti & Gaskell 1984), but this is an additional check of the orientation hypothesis because the jet axis is not necessarily lined up with the axis of rotation of the host galaxy. IC 4329A has a very steep Balmer decrement. For the radio-quiet quasars in the Stirpe (1990) sample, the Balmer decrement for the quasars showing single or double displaced peaks is indeed steeper than for those not showing such peaks[14]. The median Hα/Hβ ratios are 3.76 and 3.31 respectively. The one-tailed significance of the difference is 95%.

4. *They will be found in quasars with steep optical spectra.* A quasar's UV excess also correlates with galaxy orientation (Cheng, Danese & De Zotti 1983; De Zotti & Gaskell 1984). Again, this will be an extra check because the axis of the central engine is not necessarily aligned with the galaxy.

Except for the second of these predictions, they all obviously hold true already for radio-loud 3C 390.3 objects (the second does not hold because the radio-loud quasars are mostly in elliptical or disturbed galaxies).

## 6.8 Polarization

The polarization of the lines and continuum of quasars almost certainly occurs in scattering electrons or dust beyond the BLR. Since radio-quiet 3C 390.3 quasars are predicted to have their axes inclined to the line of sight, their broad lines should have higher polarization on average than the lines of non-3C 390.3 quasars. In both radio-loud and radio-quiet 3C 390.3 quasars, the low-velocity core will probably have a lower polarization since it is believed to come from further out. This is not a unique prediction of the SMB model. Since, in the SMB model the two peaks come from slightly different places it is possible that the polarizations will differ slightly, but this depends on the size of the scattering region. As already mentioned, Antonucci, Hurt & Agol (1995) have found significant line polarization in 3C 390.3 itself. Their results are probably consistent with the predictions of the SMB model.

## 6.9 Distribution of Extended Low Velocity Gas

In connection with the predictions of the orientation hypothesis it is worth noting that Simkin, Su & Schwarz (1980) found the FWZI of Hβ to be correlated with the inclination of the host galaxy while the FWHM of Hβ was not (see their figure 4*a*). This needs confirming with a larger data set, but the difference in the behavior of the FWZI and FWHM is easily explained if we remember that there is always an undisplaced peak due to gas surrounding the system. The undisplaced peak strongly influences the FWHM; the FWZI is free of this influence. The lack of variation of the width of the peak with inclination suggests that it has a more spherically symmetric distribution.

---

[14] One inexplicable result deserves mention. Three of the Stirpe quasars had uncertain profile classifications and so were left out of this analysis. Curiously, they all proved to have the flattest Balmer decrements! Only one other classified profile approached them in flatness of the decrement. The two-tailed significance of this is over 99.98%

Wills & Browne (1986) studied FWHM(Hβ) versus $R$. In the SMB model there are two BLR components: the broad component associated with the binary central engine and the narrower undisplaced component associated with gas further out. The SMB model predicts that the broad component will show the strongest correlation with orientation. Thus the FWZI, or something similar, will show the strongest correlation with $R$. In table 6 of Brotherton (1995) the significance of the correlation of $R_V$ (the best orientation indicator) with the full width of Hβ at the 75% level has a Spearman rank correlation coefficient of 0.45 (99.9% significance). For the full width of Hβ at the 25% level, a measure closer to the FWZI, the correlation coefficient rises to 0.61 (99.9999% significant). This strongly confirms the idea that it is the *broad base* which is orientation dependent. Brotherton et al. (1994), from a study of object-to-object variations of UV lines, identify two components: a "very broad line region" (VBLR), and an "intermediate line region" (ILR). Brotherton (1995) also identifies this component as the cause of object-to-object variations in Hβ. I believe the unshifted peak in 3C 390.3 objects is the ILR.

### 6.10 Are All Quasars Binaries?

There are certainly quasars that do not require binary central engines (quasars with single unshifted peaks and no signs of precession of their jets), but whether all quasars are binaries is a question worth exploring. The fraction of quasars showing some sort of displaced peak is already over 50% (see tables 1 and 2). Sulentic (1989) found that 50% of broad Balmer lines show a detectable displacement with respect to the rest frame of their host galaxies. My own (unpublished) study of radio-loud quasars gives an identical result. In both cases the threshold for detection of motion is about 200 km s$^{-1}$. Face-on binaries will not show double displaced peaks and, even if the system is inclined to the line of sight, there will be times in the orbit when the system looks single lined. The latter will happen for 3C 390.3 itself sometime in the years 2012 – 2029. If we make allowance for orientation effects and lower the detection threshold, the observed 50% implies that *essentially all BLRs are moving relative to their host galaxies*. This is particularly true when we consider that selection effects lead to a disproportionate number of both radio-loud and radio-quiet quasars being face-on.

There is one additional mechanism I should mention that could cause *modest* motion relative to the host galaxy without needing a binary. This is the motion of galactic cores relative to the center of mass of the galaxy. Miller & Smith (1992) discovered this in N-body simulations and they illustrate it in the movie accompanying their paper. The core of the galaxy wanders around erratically at a few hundred km s$^{-1}$ with a quasi-period of a million years or more. It is easy to move the nucleus around because the restoring force is small (S.J. Aarseth, private communication). Any nuclear motion will damp out slowly because the damping time is long (S. D. Tremaine, private communication). Many galaxies are observed to have off-center nuclei (see Miller & Smith 1992 for references). The off-center nucleus mechanism could be responsible for the small velocity shifts seen in quasars while SMBs could be responsible for the large shifts. If this is the case, I have overestimated the number of SMBs.

### 6.11 Formation and Evolution of SMBs

Since Begelman, Blandford & Rees (1980) and Roos (1981) there has been a lot of work on the formation and evolution of SMBs and the production of gravitational radiation when they merge. Discussing this is beyond the scope of this paper, but I refer the interested reader to the review article by Valtonen & Mikkola (1991) for a discussion of the "few-body problem" aspects and to more recent work by Makino et al (1993), Zamir (1993), Xu & Ostriker (1994), Makino & Ebisuzaki (1994) and Governato & Maraschi (1994). This work confirms the essential details of the Begelman, Blandford & Rees picture. Finally, Wilson & Colbert (1995) have evoked the *merger* of SMBs as a possible explanation of radio-loudness/radio-quietness. In their model, however, the *radio-loud* quasars are the ones where the massive objects have merged to produce a *single* central object with greater angular momentum. In some ways this is the opposite of the model discussed here.

### 6.12 The Binary − Jet Connection

If we consider the members of "the jet set", the various classes of objects showing jets (for overviews, see Kundt 1987, 1996), they are all binaries. This is obvious for X-ray binaries, galactic superluminals, symbiotic stars and SS 433. Consideration of double star statistics shows that proto-stellar jet sources will mostly be binaries. 53% of solar type stars in the field are binaries (Duquennoy & Mayor 1991). Close binaries will already have coalesced. This is borne out by studies of young clusters. In these there is an excess of short-period binaries (see their figure 8b). Reipurth & Zinnecker (1993) have carried out a CCD search for *visual* binaries among pre-main sequence stars. They estimate binary frequencies of 60 - 90% (see their figures 8*a,b*). Obviously the binary frequency is high in pre-main sequence stars. There are some protostellar objects where double pairs of jets are seen, but these are very wide systems (the jets would not be resolved if they were not). The general question of jet production in very close binaries needs to be considered

If quasars with jets are also binaries *then perhaps all jet sources are binaries!* Does this mean anything? We believe that galactic jet sources are binaries because we need the companion to provide the supply of material for the compact object. It is normally thought that a single central object in quasars can be fueled with no difficulty, but Roos (1981, 1988 and references therein) has discussed how a binary can enhance the fueling of a central engine in a quasar. Perhaps many (all?) quasars are binaries because a binary has a high fueling rate. For quasars, the most spectacular examples of the 3C 390.3 phenomenon have strong jets, but I have argued that this could just be an orientation effect and not necessarily anything related to jet production.

## 7. Conclusions

I have argued that both optical spectroscopy and radio observations independently give strong evidence for the existence of supermassive binaries in many or even all quasars. The main predictions of the supermassive binary picture such as radial velocity variation and orientation dependence seem to be borne out, but more work on many aspects of the picture is still needed. The production of supermassive binaries through mergers is something we know *must* have happened and, indeed, still is happening. We need to understand the consequences.

**Acknowledgments.** I would like to express my appreciation to Wolfgang Kundt for the invitation to speak at this workshop and for all the effort he put into it. The workshop and its predecessor in Erice in 1986 have been a stimulus for my thinking in a number of areas. I am grateful to Ski Antonucci for sending an advance copy of his Arp 102B polarization results. I would also like to thank Sandy Faber for some useful discussion. Partial support for this work was provided by NASA through grant number AR-05796.01-94A from the Space Telescope Science Institute, which is operated by the Association of Universities for Research in Astronomy, Inc., under NASA contract NAS5-26555.